# Revealing Quantitative Observation of Multi-Loop Super-Toroidal Currents in Plasmonic Meta-Atoms

**Burak Gerislioglu and Arash Ahmadivand[*]**

*Department of Physics and Astronomy, Rice University, 6100 Main St, Houston, Texas 77005, United States*

[*]aahmadiv@rice.edu

**ABSTRACT:** Living in a world of resonances, there have been significant progresses in the field of excitation high-quality and multifunctional moments across a wide range of optical frequencies. Among all acknowledged resonances, the toroidal multipoles have received copious interest in recent years due to having inherent signatures in nature. As a fundamental member, toroidal dipole is a strongly localized electromagnetic excitation based on charge-current distributions, which can be squeezed in an extremely tiny spot. Although there have been extensive studies on the behavior and properties of toroidal dipoles to develop all-optical devices based on this technology, so far, all analyses are restricted to the first ($1^{st}$) order toroidal dipoles. In this work, using a simple technique, we successfully monitored supporting exquisite multi-loop super-toroidal (MLST) spectral features in a planar multipixel plasmonic meta-atom. We quantitatively demonstrate that how a traditional toroidal dipole can be switched to a super-toroidal moment by varying the dielectric permittivity of the gaps between proximal pixels. This understanding paves new approaches for the excitation of complex multi-loop toroidal moments in plasmonic metamaterials with high sensitivity, applicable for various practical applications.

Charge-current configurations possess perceivable signature either in microscopic and macroscopic levels, where the electromagnetic properties of these multipole excitations, arising from any complex system, have been characterized qualitatively by multipole expansions in electrodynamics principles.[1,2] However, unconventional charge-current excitations' footprints (i.e. ring currents, poloidal magnetic fields, toroidal modes, flying doughnuts, anapoles, etc.) with concealed far-field radiation patterns, have been spotted recently in both nucleic scale and giant systems (galaxies, neutron starts, black holes, pulsars, and quasars).[3] In the subwavelength optical systems limit, the loop-currents have successfully been induced either in artificial 3D and planar all-dielectric, and plasmonic metamolecules.[4,5] Enforcing simultaneous time-reversal ($t \rightarrow -t$) and space-inversion ($r \rightarrow -r$) violations enable the excitation of significantly squeezed, gyrotropic-fashioned rotating magnetic fields in the systems, known as "toroidal multipoles".[4,6] Intensely masking by dominant classical electromagnetic far-field radiations, as ghost resonances, toroidal topology provide a nonzero contribution to the scattered radiation, written as:

$$E_s^{FF} = 4\pi k^2/c \left( \sum_{l,m} Q_{l,m}\psi_{l,m} + \sum_{l,m} M_{l,m}\phi_{l,m} + \sum_{l,m} T_{l,m}\psi_{l,m} \right) \quad ; \begin{cases} Q \propto d/dr(\mathbf{r} \cdot j_l) \\ M \propto |\mathbf{r} \times \mathbf{J}| \\ T \propto |\mathbf{r} \cdot \mathbf{J}| \end{cases} \quad (1)$$

where $Q$, $M$, and $T$ correspond to the induced charge, transversal and radial currents, respectively. Of particular interest is the dynamic toroidal dipole, the fundamental member of toroidal multipoles, has been analyzed in metamaterials and utilized extensively for developing advanced technologies and practical devices from immunosensors to modulators.[5,7,8] Such extensive interests originate from the inherent nonradiative property of toroidal dipole due to suppression of the electric dipole mode. Theoretically, the direct consequence of the combination of magnetic quadrupole and electric octupole is the toroidal dipole moment,[9] describing as: $\vec{T} = 1/10 \int_V \left\{ (\vec{\mathbf{r}} \cdot \vec{\mathbf{J}})\vec{\mathbf{r}} - 2\vec{\mathbf{r}}^2 \mathbf{J} \right\} dv$.

In terms of toroidization principle, the optically driven dynamic toroidal dipole can be imagined as magnetically induced currents fluxes across the surface of a torus, also known as poloidal currents or loop currents, along the associating meridians.[4,5,10] Traditionally, metasurfaces mimic the artificial torus arrangement for the electromagnetic excitation of the toroidal dipole spectral feature. However, the order of these toroidal currents have always been reported as "zero ($m=0$)" and "one ($m=1$)", and the excitation of toroidal modes with multiple loop-currents, or super-toroidal currents, have not been observed yet in typical metamaterial platforms. Toroidal currents with different orders ($m>1$) can be visualized as twisted doughnut-shape configurations (see Figure 1A). Depending on the number of these loop-currents, the super-toroidal current configurations can be categorized in odd and even-orders. As promising tools, numerical techniques help us to understand the possibility of the formation of multi-loop super-toroidal (MLST) current arrangements in metasurfaces. In this Report, to address this fundamental lack of knowledge in toroidal moment's theory, we report on the first quantitative observation of the excitation of MLST modes in planar plasmonic metamaterials. By tailoring a multicomponent unit cell consisting of aluminum nanopixels, we numerically computed the possibility of the excitation of unique regular toroidal dipole feature across the near-infrared spectrum. To induce the super-toroidal spectral feature, we enhanced the plasmon coupling effect between the proximal nanopixels *via* introducing alumina ($Al_2O_3$) spacers at the specific capacitive gaps. Our findings show that how the multi-loop mode can be excited and observed in plasmonic meta-atoms.

In Figure 1A, we artistically demonstrated the formation of multi-loop currents (defined by **$m_1$**, **$m_2$**, and **$m_3$**) in the proposed nanostructure, plotted based on our numerical observations. This illustration gives a brief overview for the professed assertion and formation of odd-order MLST modes. The corresponding geometrical sizes for the structure are exhibited in Figure 1B, as top-view images for vacuum condition and the presence of alumina spacers in the openings between adjacent resonators, distinguished by a different color. Under *y*-polarized beam illumination, one can see the excitation of pronounced toroidal dipole modes around $\lambda \sim 1$ µm and $\lambda \sim 1.1$ µm for the gaps in vacuum and alumina-filled regimes, respectively (Figure 1C). The transmission ($\Gamma$) spectra reveals the formation of two minima across the near-infrared band, corresponding to double directional toroidal dipoles, driven by the poloidal fields and loop-currents.[10] However, the tunable property of the nanostructure is not the concept of interest in this work. Hence, we merely consider the target wavelength for the indicated mode at shorter wavelengths for the excitation of super-toroidal feature. The distinct red-shift (around $\delta\lambda$~100 nm) in the position of toroidal dipole originates from the dielectric permittivity variations in the capacitive spots. The characteristic electric and magnetic-fields (E and H-fields) maps allows to understand the excitation of toroidal feature. The 3D |E|-field map in Figure 1D illustrates strong field confinement at the capacitive gaps (with alumina spacers), nonetheless, the intensity of the field enhancement at the opening areas close to the center of the unit cell is much stronger. This is because of the induction of an oblique toroidal mode spinning across the peripheral curved pixels, arising due to the magnetic field mismatch at these resonators. The |H|-field snapshot (in Figure 1E) confirms the realization of an oblique moment and strong confinement of magnetic fields across the curved resonators. The strong magnetic field confinement is demonstrated in the cross-sectional plane ($z$-axis) as a function of incident beam wavelength in Figure 1F. As anticipated, the contribution of the magnetic field to the scattered radiation from the scatterer at the resonance wavelength around $\lambda$~1.1 µm is much stronger. Although the excitation of oblique toroidal dipole in the proposed meta-atom is a unique feature, there is a hidden phenomenon in the structure, will be explained in the following part.

The theory for the excitation of toroidal moments with high-orders or multi-loop currents have been studied rarely and it is limited to showing the theory behind (*11*). In the dynamic toroidization physics, the induced current density (*J*) for a coaxial toroidal multi-loop currents can be written as: $J_m = T_{m-1} \nabla^{m+1} \times (\mathbf{n}\delta^m(r))$, where *T* is the toroidal dipole moment, *m* is the order of the MLST, and **n** is the vector of unit length normal to the plane of the toroidal loop. Figure 2A contains the required information for the toroidal currents pecking order, employed to define the theory and notations in the following descriptions. This profile contains the toroidal dipole from $0^{th}$ order (*m*=0) to the $3^{rd}$ order (*m*=3). Figures 2B and 2C compare the formation of the toroidal dipole moments in vacuum and the presence of alumina capacitive spacer, respectively, at and around the resonance wavelengths as vectorial H-field maps in *xz*-planes. Here, we try to monitor the quality of the toroidal spectral feature in the vicinity of the central wavelength to be sure about our claim. Clearly, for the vacuum condition, the toroidal moment is robust at $\lambda$~1 µm, and keeps its nature around the resonance position (±50 nm) and vanishes at further wavelengths. The noteworthy point here the formation of a single toroidal loop-current at all wavelengths, corresponding with the 1st order toroidal current (*m*=1). On the other hand, for the presence of alumina spacer at the gaps, we observed a distinguished toroidal loop-current at $\lambda$~1.1 µm. Besides, additional loops around the core spectral feature are noticeable, whilst it was missed in the former studies above. In this regime, the peripheral vectorial loops are correlated with the third loop-current (**m$_3$**), keeping their strength around the original feature at 100 nm away from the resonance wavelength. It should be underlined that due to showing the vectorial boards in 2D plots, it is challenging to plot all loops simultaneously. To investigate the possibility of the generation of the third toroidal current loops, one should monitor the other planes (*xy* and *yz*-planes) around the unit cell to monitor the presence of **m$_2$**. Figures 2D and 2E illustrate and evaluate the vectorial maps in *yz*-plane for the excitation of second loop-current (**m$_2$**) in the vacuum and alumina-filled gap regimes, respectively. Clearly, for the gaps without the spacer, the projected magnetic field is not perturbed, while for gaps with alumina spacer, a spinning pattern is formed correlating with the second loop (**m$_2$**). This strongly verifies the excitation of super-toroidal spectral feature in the proposed plasmonic meta-atom.

To underlay the physics behind the formation of MLST signature, one need to investigate the influence of the capacitive spacer on the spectral response of the metasurface as a key parameter. Given that the projected toroidal dipole currents strongly depend on the surrounding complex dielectric permittivity ($T_1 = -ILc^2/\omega^2\tilde{\varepsilon}$ and $T_2 = -mc^2/\omega^2\tilde{\varepsilon}$, in which $L$ is the length of the dipole and $I$ is the amplitude of the dipole loop-current), therefore, controlling the excitation of multiple toroidal components and loops could be possible by applying variations in the dielectric permittivity at the critical areas of the structure and vice versa. In other words, varying the ratio between the amplitudes of the currents flowing in the toroidal dipole can be accomplished by tuning the complex dielectric of the medium. As the gaps play fundamental role in the plasmonic response of the proposed unit cell, subtle perturbations in the dielectric permittivity of gaps alter the spectral behavior of the entire metasurface. To discover the influence of the variations in the gaps' permittivity, one should examine the relation between toroidal currents and dielectric variations. The following relations,[11,12] confirms the strong dependency of the number of loops ($N$) and currents amplitude ($I$) to the dielectric permittivity of the gap and vice versa.

$$\begin{cases} T_1 = \dfrac{N\pi R a^2 I}{8}\mathbf{n} \\ T_2 = \dfrac{NI\pi^2 R^2 \left(d^2 + 2bd\right)}{16}\mathbf{n} \end{cases} \quad ; R \ll \lambda \quad (2)$$

Ultimately, we analyzed the induced current density due to the formation of multi-loop spectral feature, as shown in Figure 3 (see Methods). This cross-sectional image ($xz$-axis) allows to compare the induced current density in the presence and absence of alumina capacitive spacer at the gap regions. Obviously, in the vacuum limit, a dominant extreme associating with the toroidal dipole feature arises ($\mathbf{m_1}$), while for the presence of alumina in the gaps, a second peak is induced corresponding to the additional loop-currents ($\mathbf{m_2}$ and $\mathbf{m_3}$), having amplitude analogous to the original peak. It should be noted that a small shoulder in the current map in the vacuum regime is formed due to the poloidal current of the 1$^{st}$ order toroidal dipole moment.

In conclusion, we have analyzed the first quantitative observation of super-toroidal feature in plasmonic metamaterials across the near-infrared region of spectrum. By developing a multipixel plasmonic nanoplatform, we have demonstrated the possibility of the excitation of MLST mode using theoretical and numerical techniques. The observation of multi-loop currents (super-toroidal mode) in planar meta-atoms was accomplished by applying perturbations in the capacitive openings between the proximal nanopixels. Such a simple and fundamental approach allowed us to enhance the surface current density *via* the excitation of high-order multi-loop toroidal features. We believe that the methodology provided here can be effectively employed for developing higher-order of toroidal moments with ultra-high sensitivity to the surrounding perturbations such as precise plasmonic immunosensors and nanophotonic devices.

# Figures Captions

**Fig. 1.** A) An artistic rendering of the proposed toroidal plasmonic meta-atom, holding multiple twisted loop-currents. B) Top-view images of the unit cell for the absence and presence of alumina spacer at the gaps with description to the geometrical sizes. C) Transmission ($\Gamma$) amplitude for the excitation of toroidal feature under polarized normal beam excitation for two different analyzed regimes. D) The 3D |E|-field map for the excitation of toroidal feature in the presence of alumina spacer at the resonance wavelength ($\lambda \sim 1100$ nm). E) The 2D |H|-field map for the excitation and localization of the magnetic field, leading to inducing oblique toroidal feature. H) The cross-sectional |H|-field snapshot as a function of incident beam, validating emission of huge field from the toroidal scatterer at the resonance wavelength.

**Fig. 2.** A) The four members of the toroidal dipole with different orders from magnetic current-loop to $3^{rd}$ order super-toroidal feature. B, C) Vectorial H-field boards for the formation of $1^{st}$ order toroidal and super-toroidal features in *xz*-planes, respectively. In (C) the appeared peripheral loops around the central feature (**m₁**) are correlating with $m_3$. D, E) Vectorial H-field maps in yz-planes for the possibility of the excitation of the targeted loop (**m₂**) to verify the formation of third loop, in vacuum and alumina-filled gap regimes, respectively.

**Fig. 3.** The characteristic surface current density map, calculated for vacuum and alumina-filled gap regimes, showing the contribution of the super-toroidal mode in the formation of second current density extreme.

Fig. 1

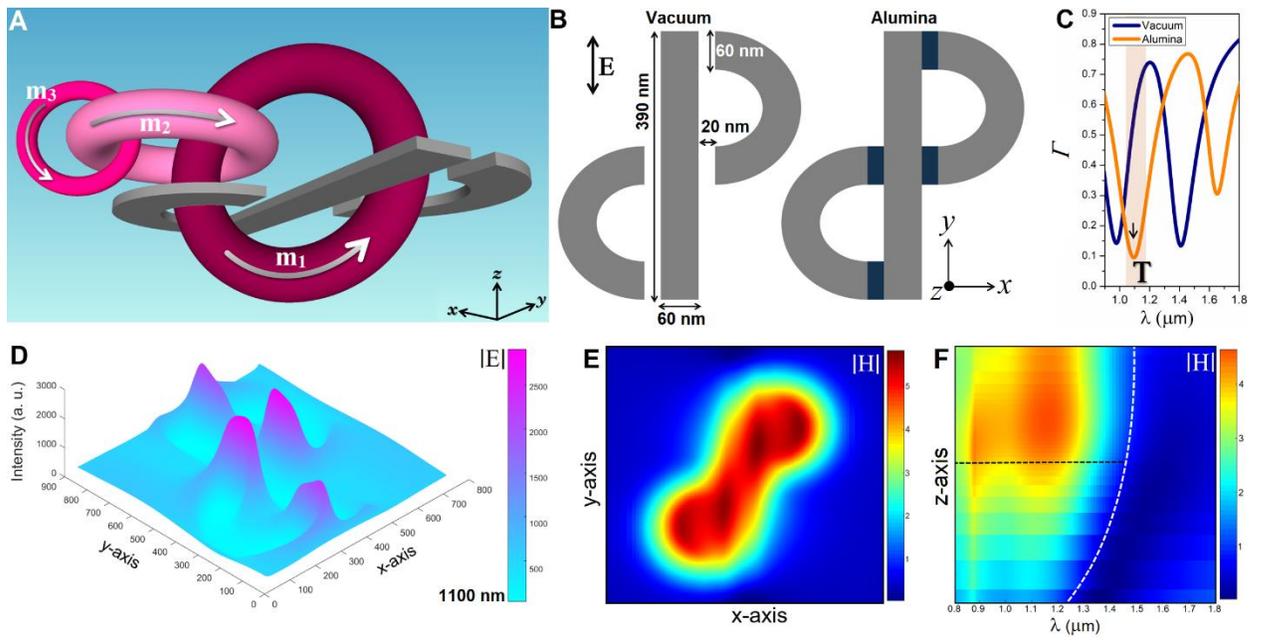

Fig. 2

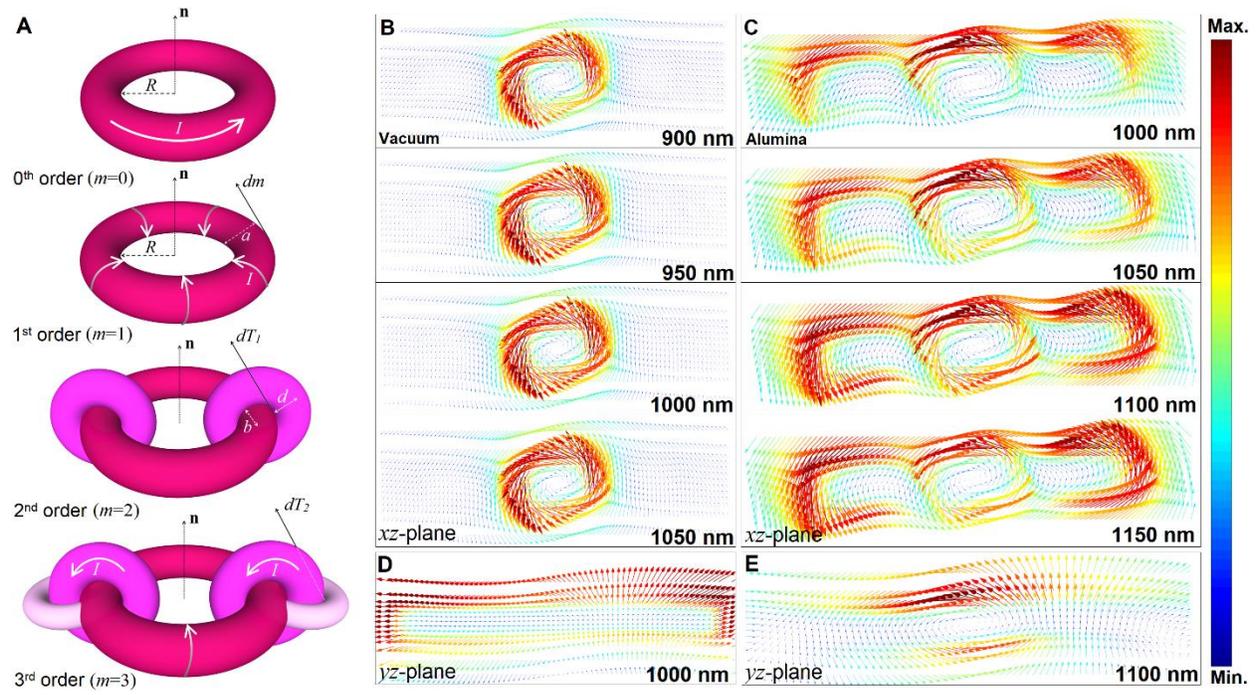

Fig. 3

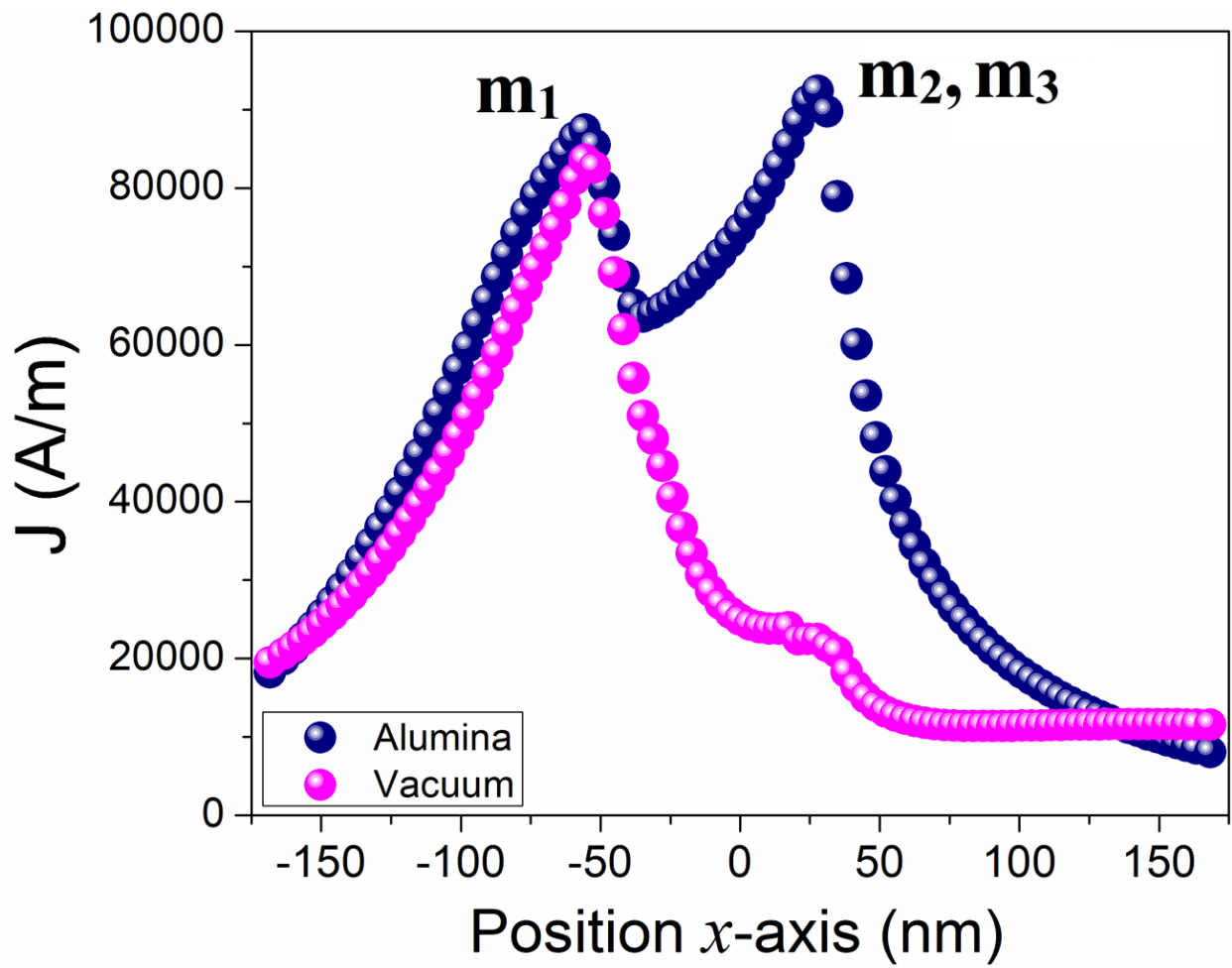